\documentclass[11pt]{article}
\usepackage{amsmath,amssymb,amsfonts}
\usepackage{units}
\usepackage{bbold,palatino}
\usepackage{graphicx}
\usepackage{dcolumn}
\usepackage{bm,bbm,mathtools,esvect}
\usepackage{slashed}
\usepackage{esvect}
\usepackage{fullpage}
\usepackage[utf8]{inputenc}
\title{\bf DEFORMED ANGULAR MOMENTUM ALGEBRA\\ WITHIN THE REAL HILBERT SPACE}

\author{{\sf SERGIO GIARDINO\footnote{\tt sergio.giardino@ufrgs.br}}\\
\\
\small \it Departamento de Matem\'atica Pura e Aplicada \\
\small \it Universidade Federal do Rio Grande do Sul (UFRGS)\\
\small \it Caixa Postal 15080, 91501-970  Porto Alegre RS \\
\small \it Brazil}
\begin{document}
\date{} 
\maketitle

\begin{abstract}
\noindent Starting from generalized position operators, we derive complex and quaternionic angular momentum operators along with their commutation algebra as well. These algebras differ from the standard Hermitian ones, especially in terms of commutation relations involving partial and total angular momentum operators. Despite these differences, the effective quantum expectation values obtained from slightly deformed algebras align with those from the conventional Hermitian algebra. This suggests that even though the wave functions and resulting dynamics differ from standard quantum Hermitian behavior, these deformed algebras can still be effectively understood as valid angular momentum algebras.

\vspace{2mm}

\noindent {\bf keywords:} quantum mechanics; formalism; other topics in mathematical methods in physics

\vspace{1mm}

\noindent {\bf pacs numbers:} 03.65.-w; 03.65.Ca; 02.90.+p.
\end{abstract}

\vspace{1cm}


\hrule
\tableofcontents
\vspace{1cm}
\hrule
\pagebreak
\section{ INTRODUCTION\label{OI}}
  
In this article, we revisit the important topic of quantum angular momentum, an extensively studied subject that already inspired several textbooks concerning it \cite{Edmonds:1955fi,Biedenharn:1981er,Varshalovich:1988ifq,Chaichian:1998kd,Devanathan:2002aam}, and continues to motivate contemporary research work on it. By way of example, current developments in quantum angular momentum encompass algebraic applications  \cite{Bhandari:2024msa}, curved space	 effects \cite{Burgarth:2025kxm,Palmerduca:2025vvd}, atomic emission \cite{Berman:2025iom}, inertial frames of reference \cite{daSilva:2025etb}, large angular momenta \cite{Konishi:2024bzf}, algebraic methods \cite{Friedmann:2012tr}, quantum optics \cite{Babiker:2013aml,Padgett:2017epa}, field theories \cite{Palmerduca:2025vvd,Lorce:2015ela}, gravitation \cite{Bhandari:2024msa,Mao:2023evc,Heissenberg:2024umh}, the Wigner-Vlasov formalism \cite{Perepelkin:2025izs}, and much further work.

Within the ambit of the real Hilbert space ($\mathbbm R$HS) formulation of quantum mechanics encoded in \cite{Giardino:2018lem,Giardino:2018rhs}, in this article we consider the particular problem of the quantum angular momentum. Remembering an essential element of this approach to be the expectation value of a physical quantity that is encoded in the operator $\widehat{\mathcal O}$ that acts over the wave function $\Psi$ to be the real quantity
\begin{equation}\label{int01}
	\big<\mathcal O\,\big>=\int\Big\{\Psi,\,\widehat{\mathcal O}\,\Psi\Big\} d\bm x,
\end{equation}
where the curly bracket inner product encompasses the probability density defined as
\begin{equation}
	\Big\{\Psi,\,\widehat{\mathcal O}\Psi\Big\}=\frac{1}{2}\left[\Psi^\dagger\widehat{\mathcal O}\Psi+\Psi\big(\widehat{\mathcal O}\Psi\big)^\dagger\right]
\end{equation}
where $\Psi^\dagger$ is of course the conjugate of $\Psi$. The expectation value (\ref{int01}) is evaluated over the real numbers, and consequently the real Hilbert space formulation  does not constrain $\widehat{\mathcal O}$ to be an Hermitian operator. Moreover, the wave function $\Psi$ takes values in quaternions, and consequently complex and real wave functions comprise particular cases. Compared to the usual complex Hilbert space formulation of quantum mechanics ($\mathbbm C$QM), the $\mathbbm R$HS approach is evidently more general and less restrictive  because the operators and wave functions are mathematically broader in scope. Therefore, in this article we examine a generalized proposal for the angular momentum operator that replaces the conventional $\mathbbm C$QM definition.

Besides discussing the quantum angular momentum operator, the research reported in this paper contributes to the broader investigation of the generality of quantum mechanics, where the real Hilbert space formalism was initially thought to be way to  solve the breakdown of the classical limit within the anti-Hermitian construction of quaternionic quantum mechanics ($\mathbbm H$QM) ({\em cf.} Sec. 4.4 of \cite{Adler:1995qqm}). We remember the introduction of the real Hilbert space formalism \cite{Giardino:2018lem,Giardino:2018rhs} as a suitable way to solve several unanswered open problems in $\mathbbm H$QM, such as the geometric phase \cite{Giardino:2016xap}, the free particle \cite{Giardino:2017yke,Giardino:2017pqq,Giardino:2024tvp}, the Virial's theorem \cite{Giardino:2019xwm,Giardino:2025bym}, the rectangular potential \cite{Giardino:2020cee,Giardino:2026buh}, the quantum  scattering \cite{Giardino:2020ztf,Hasan:2019ipt,Giardino:2025xth}, the harmonic oscillator \cite{Giardino:2021ofo},  spin \cite{Giardino:2023spz}, and the generalized wave equation \cite{Giardino:2023uzp}. Moreover, the real Hilbert space is also well suited to relativistic models \cite{Giardino:2021lov,Giardino:2021mjj,Rosa:2025git}, including field quantization \cite{Giardino:2022kxk,Giardino:2022gqn}.

One also recognizes this article as an element of the analysis of the viability of real quantum mechanics ($\mathbbm R$QM), with reported advantages \cite{Finkelstei:2022rqm,Chiribella:2022dgr,Zhu:2020iml,Fuchs:2022rih,Vedral:2023pij}, as well as drawbacks \cite{Renou:2021dvp,Chen:2021ril,Wu:2022vvi}. Importantly, although the real Hilbert space presented here is not restricted to real wave functions, thereby consisting of a unifying theory. Another key aspect of this discussion concerns the complex nature of quantum mechanics, covering topics such as quantization methods \cite{Prvanovic:2020clx}, mathematical foundations \cite{Takatsuka:2025piu,Volovich:2025rmi,Hita:2025okv}, pure imaginary elements \cite{Gour:2018qiq,Xu:2023xdb,Wu:2023lyi,Zhang:2025jdb}, locality \cite{Feng:2025eci}, unitary properties \cite{Fernandes:2024sik}, complex scalar potentials \cite{Khantoul:2022bam,Cannata:2012uc}, quaternionic Hilbert spaces \cite{Nyirahafashimana:2025rho}, open systems \cite{Acevedo:2025juf}, among others. As directions for future research, several of these complex formulations can be examined through the lens the real Hilbert space approach considered in this article, implying the generalizing requirement that the $\mathbbm R$QM theory considered here is expected to fulfill.

\section{COMPLEX SOLUTION \label{CAM}}

The standard quantum theory of angular momentum begins after defining the usual classical definition in terms of quantum operators, and hence the operator  $\widehat{\bm l}$ of angular momentum is defined in terms of the vector product
\begin{equation}\label{cam03}
	\widehat{\bm l}=\widehat{\bm r}\bm\times\widehat{\bm p},
\end{equation}
where  the position operator $\widehat{\bm r}$ reads
\begin{equation}
\widehat{\bm r}=\big(x,\,y,\,z\big)
\end{equation}
and to the linear momentum operator $\widehat{\bm p}$ corresponds 
\begin{equation}
	\widehat{\bm p}=-i\hbar\bm\nabla.
\end{equation}
In the $\mathbbm R$HS approach, the position operator $\widehat{\bm r}$ admits a generalization in complex terms \cite{Giardino:2025bym}, such as
\begin{equation}\label{cam12}
\widehat{\bm z}=\bm r+i\,\bm s
\end{equation}
where 
\begin{equation}
	\bm s=\bm s\big(\bm r\big)
\end{equation}
is an arbitrary real vector function. As discussed in \cite{Giardino:2025bym} the position expectation value calculated from (\ref{int01}) does not change after the complex generalization, so that
\begin{equation}
	\big<\bm z\big>=\big<\bm r\big>.
\end{equation}
This fact permits to interpret the vector function $\bm s$ to act as a gauge potential function that does not change the observed physical property. Therefore, we define the complex linear momentum operator to be
\begin{equation}\label{cam01}
	\widehat{\bm \ell}=\widehat{\bm z}\bm\times\widehat{\bm p}.
\end{equation}
Using an eigenfunction of $\widehat{\bm l}$, we accordingly obtain
\begin{equation}
\big<\bm \ell\big>=\big<\bm l\big>,
\end{equation}
thus emphasizing the generalized character of $\widehat{\bm\ell}$. We may thus obtain the commutator
\begin{equation}\label{cam13}
	\left[\,\widehat \ell_a,\,\widehat \ell_b\,\right]=\epsilon_{abc}\,i\,\hbar\,\widehat \ell_c+i\widehat h_{ab},
\end{equation} 
where $\epsilon_{abc}$ is the well known Levi-Cività anti-symmetric symbol, and the deformation factor $\widehat h_{ab}$ of the algebra reads
\begin{equation}
	\widehat h_{ab}=\Big[\Big(\widehat \ell_a\,\epsilon_{bmn}-\widehat \ell_b\,\epsilon_{amn}\Big)s_m\Big]\widehat p_n,
\end{equation}
or else
\begin{equation}\label{cam15}
	\widehat h_{ab}=\epsilon_{abc}\,\Big[\,\Big(\widehat{\bm\ell}\, s_c\Big)\bm\cdot\widehat{\bm p}-\Big(\widehat{\bm\ell} \bm{\cdot s}\Big)\widehat p_c\,\Big].
\end{equation}
In order to examine the consequences of the definition (\ref{cam01}), we propose
\begin{equation}\label{cam05}
	\bm s=\big(\epsilon_1 x,\,\epsilon_2 y,\,\epsilon_3 z\big)
\end{equation}
where $\epsilon_a$ are real constants and $a=\{1,\,2,\,3\}$.  In order to determine the deformation factor (\ref{cam15}), one immediately observes that 
\begin{equation}
\widehat{\bm\ell} \bm{\cdot s}=\bm 0
\end{equation}
and the nonzero contribution to comes from
\begin{equation}
\Big(\widehat{\bm\ell}\, s_c\Big)\bm\cdot\widehat{\bm p}=\epsilon_{kmn} r_m\big(\widehat p_n\epsilon_cr_c\big)\widehat p_k=-i\hbar\epsilon_c\big(\epsilon_{kmc}r_m\widehat p_k\big)=i\hbar \epsilon_c\widehat l_c
\end{equation}
where repeated indices are summed. Consequently obtaining the commutation algebra (\ref{cam13}) as
\begin{equation}\label{cam02}
	\left[\,\widehat \ell_a,\,\widehat \ell_b\,\right]=\epsilon_{abc}\,\hbar\,\big(i-\epsilon_c\big)\,\widehat \ell_c,
\end{equation}
As expected, the usual angular momentum commutator algebra is obtained within the limit $\epsilon_a\to 0$. Nonetheless, one may understand (\ref{cam02}) as a deformation of the usual angular momentum algebra generated by the components of (\ref{cam03}). For the time being, it seems not clear the precise classification of this deformed algebra, whether quantum algebraic \cite{Monteiro:1993gp}, non-commutative \cite{Eriksen:2017ndt}, or some other possibility. In these possibilities, the elements of the deformed algebra belong to a larger structure that is usually built in terms of a Hopf algebra. By way of example, it has already been determined that non-associative structures may rise up from the coexistence of quaternion fields in a non-commutative space-time \cite{Giardino:2012ti}, and further richer structure may also be found if the deformed linear momentum are elements of a larger algebraic structure like a group algebra or an universal enveloping algebra. The investigation of this matter constitutes an interesting direction for future research, but what we can immediately do is to examine the consequences generated in virtue of the deformation. Defining the total angular momentum as
\begin{equation}
	\ell^2=\bm{\ell\cdot\ell},
\end{equation}
we obtain
\begin{equation}\label{cam09}
\left[\,\widehat \ell^{\;2},\,\widehat \ell_a\,\right]=\hbar\big(\epsilon_c-\epsilon_b\big)\,\epsilon_{abc}\Big(\,\widehat\ell_b\,\widehat\ell_c+\widehat\ell_c\,\widehat\ell_b\,\Big),
\end{equation}
but also registering the general result obtained from (\ref{cam13}) to be
\begin{eqnarray}
	\left[\,\widehat \ell_a^{\;2}+\widehat \ell_b^{\;2}+\widehat \ell_c^{\;2},\,\widehat \ell_a\,\right]\!\!\!&=&\!\!\! i\left(\widehat\ell_b\,\widehat h_{ba}+\widehat h_{ba}\,\widehat\ell_b+\widehat\ell_c\,\widehat h_{ca}+\widehat h_{ca}\,\widehat\ell_c\right).
\end{eqnarray}
Therefore, $\ell^2$ does not commute with the angular momentum components, and it is not the Casimir element of the algebra, contrarily to that observed in the standard non-deformed case. Irrespective of this, imposing
\begin{equation}\label{cam08}
\epsilon_1=\epsilon_2=\varepsilon\qquad\mbox{and}\qquad \epsilon_3=0,
\end{equation}
on (\ref{cam05}), we build the operators
\begin{equation}
	\widehat\ell_\pm=\widehat\ell_1\pm i\,\widehat\ell_2,
\end{equation}
so that
\begin{equation}\label{multi87}
	\widehat\ell_+=\widehat l_+ -i\varepsilon\widehat\lambda_+\qquad\mbox{e}\qquad 
	\widehat\ell_-=\widehat l_- +i\varepsilon\widehat\lambda_-,
\end{equation}
where  of course
\begin{equation}
	\widehat l_\pm=\widehat l_1\pm i\,\widehat l_2,
\end{equation}
are the usual ladder operators obtained in the standard non deformed version of the quantum angular momentum theory, and
\begin{equation}
	\widehat\lambda_\pm=e^{\pm i\phi}\sin\theta\left(r\cos\theta\frac{\partial}{\partial r}-\sin\theta\frac{\partial}{\partial\theta}\right).
\end{equation}
In order to determine the influence of the deformation in terms of eigenvalues and eigenfunctions of the angular momentum operators, we initially consider the eigenfunction problem, where the total angular momentum operator
\begin{equation}
	\widehat\ell^{\;2}=\widehat\ell_+\widehat\ell_-+\widehat\ell^{\;2}_3-\hbar\,\widehat\ell_3
\end{equation}
assumes the expression
\begin{equation}\label{cam04}
	\widehat\ell^{\;2}=\widehat l_+\widehat l_- +i\varepsilon\Big(\,\widehat l_+\widehat\lambda_- - \widehat\lambda_+\widehat l_-\Big)+\varepsilon^2\,\widehat\lambda_+\widehat\lambda_- +\widehat\ell^{\;2}_3-\hbar\,\widehat\ell_3
\end{equation}
In terms of a weak deformation, we assume
\begin{equation}
	\varepsilon\ll 1,
\end{equation}
and thus authorizing to discard the term of (\ref{cam04}) that depends on $\varepsilon^2$. Proposing the wave function
\begin{equation}
	\psi=r^k e^{m(i+\varepsilon)\phi}f(\theta),
\end{equation}
where $f$ is a function to be determined, we obtain
\begin{equation}\label{multi86}
	\widehat\ell^{\;2}\psi=\hbar^2r^ke^{m(i+\varepsilon)\phi}\Bigg[\Big(1+2\varepsilon\sin^2\theta\Big)\widehat\alpha+\varepsilon \widehat\beta\Bigg]f,
\end{equation}
where 
\begin{equation}
	\widehat\alpha=-\frac{d^2}{d\theta^2}-\cot\theta\frac{d}{d\theta}+\frac{m^2}{\sin^2\theta}
\end{equation}
and
\begin{equation}
	\widehat\beta=\big(2k+5\big)\sin\theta\cos\theta\frac{d}{d\theta}\,+\,k\big(1-3\cos^2\theta\big)+m\big(2\cot^2\theta-1\big)-2m^2.
\end{equation}
Remembering
\begin{equation}
	\widehat \alpha\, P_{\,\lambda}^{\,m}=\lambda\big(\lambda+1\big)P_{\,\lambda}^{\,m},
\end{equation}
where $P_{\,\lambda}^{\,m}(\theta)$ are associate Lagrange polynomials and to the $\lambda$ eigenvalue corresponds integer and semi-integer numbers, we announce the function
\begin{equation}
	f(\theta)=P_{\,\lambda}^{\,m}(\theta)+\varepsilon Q(\theta),
\end{equation}
where $P_{\,l}^{\,m}(\theta)$ is the usual associate Lagrange polynomial that must be recovered
in the non deformed limit, and $Q(\theta)$ has to be determined from 
\begin{equation}\label{cam07}
	\Big[\widehat\alpha-\lambda\big(\lambda+1\big)\Big] Q=\Big[\kappa-2\lambda\big(\lambda+1\big)\sin^2\theta-\widehat\beta\Big] P_{\,l}^{\,m}.
\end{equation}
According to the usual perturbation theory, it is possible to to obtain approximate wave functions using the expansion
\begin{equation}\label{cam06}
	Q=\sum_{k=-m}^m C_k^{(m)}P^m_k.
\end{equation}
Then, we substitute (\ref{cam06}) in (\ref{cam07}), and picking  $k=\lambda$ the right hand of \ref{cam07} disappears, and after the inner product with $P_{\,l}^{\,m}$ determines $\kappa$. On the other hand, whenever $k\neq\lambda$, we obtains the $C_k^{(m)}$ coefficients after the inner product with $P_{\,l}^{\,m}$, exactly following the perturbation theory method.

After determining the general deformed algebra (\ref{cam02}) and the wave functions within the ambit of the (\ref{cam08}) choice, we examine whether the deformed angular momentum operators still conserve the physical interpretation of the usual non deformed case.

As a last point, one has to determine whether the deformed operators $\widehat\ell^{\,2}$, $\widehat\ell_3$ and $\widehat\ell_\pm$ have similar physical interpretations to the corresponding operators of the non-deformed complex case. Using (\ref{cam02}-\ref{cam09}) and $\epsilon_3\neq 0$ we obtain the commutators
\begin{equation}\label{cam11}
	\Big[\,\widehat\ell_3,\,\widehat\ell_\pm\Big]= \pm\hbar\big(1+ i\varepsilon\big)\,\widehat\ell_\pm, \qquad \qquad\left[\,\widehat \ell^{\;2},\,\widehat \ell_3\,\right]=0,
\end{equation}
\begin{equation}\label{cam10}
	\Big[\widehat\ell^{\;2},\,\widehat\ell_\pm\Big]=i\,\hbar\big(\varepsilon-\epsilon_3\big)\Big\{\widehat\ell_\pm,\,\widehat\ell_3\Big\},
\end{equation}
as well as the relation
\begin{equation}
	\widehat\ell_\pm\widehat\ell_\mp=\widehat\ell^{\,2}-\widehat\ell^{\,2}_3\pm\hbar\big(1+i\epsilon_3\big)\widehat\ell_3.
\end{equation}
Repeating the choice $\epsilon_3=0$, and putting into service (\ref{cam11}) and the relation
\begin{equation}\label{multi88}
	\widehat\ell_3\widehat\ell_\pm=\Big[\widehat\ell_3,\,\widehat\ell_\pm\Big]+\widehat\ell_\pm\widehat\ell_3,
\end{equation}
we obtain that the wave function obtained from $\widehat\ell_\pm\psi$ is eigenfunction of the operator $\widehat\ell_3$ with the eigenvalue
\begin{equation}
	\hbar\Big[m\pm\big(1+i\varepsilon)\Big],
\end{equation}
and thus $\widehat\ell_\pm$ are still interpreted as rising and lowering operators despite the deformation in the algebra. On the other hand, this change does not change the expectation value associated to this eigenvalue, because the imaginary component does not contribute. Consequently, the physical understanding of the $\widehat\ell_\pm$ is kept within the deformed framework. On the other and, the wave functions  $\widehat\ell_\pm\psi$ are not eigen-functions of $\widehat\ell^{\,2}$ because of the non-commutativity observed in (\ref{cam10}). However, considering 
\begin{equation}
	\widehat\ell^{\,2}\widehat\ell_\pm=\Big[\widehat\ell^{\;2},\,\widehat\ell_\pm\Big]+\widehat\ell_\pm\widehat\ell^{\;2},
\end{equation}
we effectively assume
\begin{equation}
	\left<\Big[\,\widehat\ell^{\;2},\,\widehat\ell_\pm\Big]\right>=0,
\end{equation}
because of the dependence on imaginary terms and higher order $\varepsilon$ terms. Therefore, we can still understand the deformation as keeping the observed physical behavior of quantum angular momentum operator, despite the deformation. Further understanding of the consequences of the deformed algebra are exciting direction for future research.
\section{QUATERNIONIC SOLUTIONS \label{CSO}}
In this section, our aim is to show that the deformation observed in the complex case is also observed within quaternionic cases. The approach here is however descriptive, because  the quaternionic cases are algebraically much more complicated, and additional physical novelties are not expected. Before starting, we remember that quaternionic wave functions require two complex components, respectively $\psi_0$ and $\psi_1$, so that
\begin{equation}\label{ci20}
	\Psi=\psi_0+\psi_1 j,
\end{equation}
where $j$ corresponds the quaternionic imaginary unit. Fundamentals of quaternion theory are encountered in several sources \cite{Ward:1997qcn,Morais:2014rqc,Ebbinghaus:1990zah} and will not be provided here. We only recall the anti-commutativity property
\begin{equation}
	ij=-ji,
\end{equation}
that determines
\begin{equation}
	\Psi i\neq i\Psi.
\end{equation}
Because of this, two equations of motion are possible, firstly the left wave equation
\begin{equation}
	i\hbar\frac{\partial\Psi}{\partial t}=\widehat{\mathcal H}_L\Psi,
\end{equation}
and also the right wave equation
\begin{equation}
	\hbar\frac{\partial\Psi}{\partial t}i=\widehat{\mathcal H}_R\Psi.
\end{equation}
The left and right Hamiltonian operators have particular linear momentum operators, whose action over a quaternionic arbitrary wave function $\Psi$ respectively read
\begin{equation}
	\widehat p_L \Psi=-i\hbar\bm\nabla\Psi 
\end{equation}
and
\begin{equation}
	\widehat p_R \Psi=-\hbar\bm\nabla\Psi i
\end{equation}
Several applications of these linear momenta have already been considered \cite{Giardino:2024tvp,Giardino:2025bym,Giardino:2026buh}, and before examining both of the cases, we remember the quaternionic position operator $\widehat{\bm q}$ defined as
\begin{equation}
\widehat{\bm q}=\bm z+\bm wj
\end{equation}
where the complex generalized position has been defined in (\ref{cam12}), and
\begin{equation}
	\bm w=\bm w\big(\bm r\big)
\end{equation}
is an arbitrary complex vector function. After these definitions, we finally inspect each case separately.
\subsection{LEFT QUATERNIONIC CASE}
In this case the left angular momentum will be
\begin{equation}
	\widehat{\bm\ell}_L=\bm{q\times p}_L.
\end{equation}
or else
\begin{equation}
	\widehat{\bm\ell}_L=\widehat{\bm\ell}+\hbar\,i\,j\,\bm{w^\dagger\!\times\!\nabla}.
\end{equation}
The deformed angular momentum algebra becomes
\begin{equation}
	\Big[\widehat\ell_{L\,a},\,\widehat\ell_{L\,b}\Big]=\hbar\,\epsilon_{abc}\,
	\widehat\ell_{L\,c}i+\widehat h_{L\,ab},
\end{equation}
and the deformation term reads
\begin{equation}\label{cam14}
	\widehat h_{L\,ab}=\hbar^2\epsilon_{amn}\epsilon_{bkl} \left[q_m \partial_n\frac{i\big(q_k-\overline q_k\big)i}{2}\partial_l-q_k \partial_l\frac{i\big(q_m-\overline q_m\big)i}{2}\partial_n +\big(q_miq_k-q_kiq_m\big)i\,\partial_n\partial_l\right].
\end{equation}
Moreover,
\begin{eqnarray}\nonumber
	\left[\,\widehat \ell^{\,2}_L,\,\widehat \ell_{L\,a}\,\right]
	\!\!\!&=&\!\!\! 
	\hbar\,\epsilon_{abc}\Big[\ell_{L\,b}\big(i\ell_{L\,c}-\ell_{L\,c}i\big)+\ell_{L\,c}\big(i\ell_{L\,b}-\ell_{L\,b}i\big)\Big]+\\
	\label{multi95}&+&\!\!\!\widehat\ell_{L\,b}\,\widehat h_{L\,ba}+\widehat h_{L\,ba}\,\widehat\ell_{L\,b}+\widehat\ell_{L\,c}\,\widehat h_{L\,ca}+\widehat h_{L\,ca}\,\widehat\ell_{L\,c}.
\end{eqnarray}
In fact, the deformation (\ref{cam14}) does not present novel physical elements compared to the previous case, because the deformation term (\ref{cam14}) is more complicated compared to (\ref{cam02}). The deformation generated by this term propagates to the further commutation relations, such as that verified in (\ref{cam11}-\ref{cam10}), and therefore the interpretation of the results also follows the complex case.

\subsection{RIGHT QUATERNIONIC CASE}

In this final situation, the right angular momentum reads
\begin{equation}
	\widehat{\bm\ell}_R=\bm{q\times p}_R.
\end{equation}
and consequently the operator algebra will be
\begin{equation}
	\Big[\widehat\ell_{R\,a},\,\widehat\ell_{R\,b}\Big]=\hbar\,\epsilon_{abc}\Big(
	\widehat\ell_{R\,c}\,\Big|i\Big)+\widehat h_{R\,ab}.
\end{equation}
Defining the notation for $a,\,b,\,f\in\mathbbm H$,
\begin{equation}
(a|b)f=afb
\end{equation}
we notice the deformation factor to obey
\begin{equation}
	\widehat h_{R\,ab}=\hbar^2\epsilon_{amn}\epsilon_{bkl}\Bigg[q_k\partial_l\frac{q_m-\overline q_m}{2}\partial_n-q_m\partial_n\frac{q_k-\overline q_k}{2}\partial_l+\big(q_mq_k-q_kq_m\big)\partial_n\partial_l\Bigg].
\end{equation}
And finally 
\begin{eqnarray}\nonumber
	\left[\,\widehat \ell^{\;2},\,\widehat \ell_{R\,a}\,\right]\!\!\!&=&\!\!\!\widehat\ell_{R\,b}\,\widehat h_{R\,ba}+\widehat h_{R\,ba}\,\widehat\ell_{R\,b}+\widehat\ell_{R\,c}\,\widehat h_{R\,ca}+\widehat h_{R\,ca}\,\widehat\ell_{R\,c}.
\end{eqnarray}	
Although simpler than the left quaternionic case, the physical situation is quite similar, and the only consequence is the form of the deformation term, while the physical interpretation is conserved.

\section{CONCLUSION\label{OCO}}

The generalized formulation of the quantum angular momentum theory presented in this paper provides further evidence supporting the consistency of $\mathbbm H$QM within the $\mathbbm R$HS as a generalized quantum theory. The commutation algebra of complex and quaternionic angular momenta has been determined, revealing that a deformation appears in all cases. Contraritly to the standard quantum theory, the deformation eliminates the possibility of a common eigenfunction for the angular momentum and the total angular momentum operators. However, in terms of expectation values, the structure of the angular momentum is effectively preserved.

In summary, this article presents a novel, general, and consistent way to deform the angular momentum algebra, in the same fashion the commutator between position and momentum has been deformed in \cite{Giardino:2025bym}. The result thus confirms the general feature of quaternionic and complex formulations to deform quantum algebras, as well as presents a simple way to generate such deformations. Therefore, various directions of future research are possible, involving quantum algebras \cite{Monteiro:1993gp}, non-abelian algebras \cite{Eriksen:2017ndt}, Trotter decomposition  \cite{Facchi:2025zje} and of course spin.

\begin{footnotesize}
\paragraph{FUNDING} The author gratefully thanks for the financial support by Fapergs under the grant 23/2551-0000935-8 within Edital 14/2022.

\paragraph{DATA AVAILABILITY STATEMENT} The author declares that data sharing is not applicable to this article as no data sets were generated or analyzed during the current study.

\paragraph{DECLARATION OF INTEREST STATEMENT} The author declares that he has no known competing financial interests or personal relationships that
could have appeared to influence the work reported in this paper.
\end{footnotesize}
%
%
%
%
\begin{footnotesize}

\end{footnotesize}
\end{document}